\documentclass[12pt, a4paper]{article}
\usepackage{amsfonts,amssymb,amsmath,amsthm,graphicx, slashed, epsfig}

\newcommand{\be}{\begin{equation}}
\newcommand{\ee}{\end{equation}}
\newcommand{\bea}{\begin{eqnarray}}
\newcommand{\eea}{\end{eqnarray}}
\newcommand{\beq}{\begin{eqnarray}}
\newcommand{\eeq}{\end{eqnarray}}

\def\({\left(}
\def\){\right)}
\def\[{\left[}
\def\]{\right]}
\def\a{\alpha}

\def\bp{{\mathbf p}}
\def\bq{{\mathbf q}}

\def\tr{{\rm tr}}

\begin{document}

\title{Theory of the Casimir effect for
graphene at finite temperature}

\author{Valery N. Marachevsky \thanks{email: maraval@mail.ru} \\
{\it Department of Theoretical Physics} \\ {\it Saint-Petersburg
State University}\\
{\it 198504 St. Petersburg, Russia.} }

\date{}

\maketitle

\begin{abstract}
Theory of the Casimir effect for a flat graphene layer interacting
with a parallel flat material is presented in detail.
 The high-temperature asymptotics of a free energy in a graphene-metal
 system coincides with a Drude high-temperature
asymptotics of the metal-metal system. High-temperature behavior in
the graphene-metal system is expected at separations of the order of
$100$ nm at temperature $T=300$K.

\end{abstract}


\section{Introduction}
The quasiparticles in graphene\cite{Geim} obey a linear dispersion
law $\omega= v_F k$ ($v_F\approx c/300$ is a Fermi velocity, $c$ is
a speed of light) at energies less than $2$ eV. Graphene's $2+1$ -
dimensionality and quasi-relativistic Dirac model for its
quasiparticles make it possible to derive the Casimir effect
properties of graphene systems from general constraints and
principles of quantum field theory.

Casimir effect in graphene systems was studied in different papers
\cite{Dobson}-\cite{Ali}. Finite-temperature results were obtained
in Refs. \cite{Gomez} and \cite{Mar1}. In the current paper we
follow the formalism developed in Ref.\cite{Mar1}, an alternative
derivation of the reflection coefficients is presented, the relation
to Feynman diagrams is discussed. First we derive the expressions
for the components of the polarization operator of quasiparticles in
a graphene layer at finite temperature and derive the reflection
coefficients of a flat graphene layer from the solutions of the
boundary problems for vector potentials. The free energy is given
then in two equivalent forms: in terms of reflection coefficients
and closed Feynman diagrams. Finally we study the exceptional
properties of the free energy of a flat graphene layer -- parallel
flat metal system at finite temperatures.

We use the coordinates $x^3$ and $z$ interchangeably throughout the
paper. When needed we select the coordinate $y$ along a direction of
the wavevector $\bq=(q^1,q^2)$ (longitudinal direction) and the
coordinate $x$ along a transverse direction.
 We use
$\hbar=c=k_B=1$.

\section{Action and polarization operator}
The model is described by the following classical action (assuming
graphene plane lying at $x^3=0$)
\begin{equation}
    S= -\frac14 \int d^4x\, F^2_{\mu\nu}+
        \int d^3x
        \bar\psi \slashed{D} \psi
    \label{action}
\end{equation} with
\begin{equation*}
     \slashed{D} = (i\partial_0-\mu-eA_0)\gamma_0
        +v_F[\gamma^1\(i\partial_1-e{A}_1\) +\gamma^2\(
i\partial_2-eA_2\)] -m\,.
\end{equation*}
 Here $\mu$ is a
chemical potential, $m$ is a mass gap of quasiparticle excitations,
$v_F\approx 1/300$ is a Fermi velocity. Since there are $N=4$
species of fermions in graphene, the gamma matrices are in fact
$8\times 8$, being a direct sum of four $2\times 2$ representations
(with two copies of each of the two inequivalent ones),
$\gamma_0^2=-(\gamma^{1,2})^2=1$. The Maxwell action is normalized
in such a way that
\begin{equation}
    e^2\equiv 4\pi\alpha =\frac{4\pi}{137}.
    \label{e}
\end{equation}

In Minkowski space the one-loop polarization operator can be
expressed in momentum space as
\begin{equation} \Pi^{mn}(p_0,{\bf p})
    =ie^2 \int\frac{dq_0d^2 {\bf q}}{(2\pi)^3}\,\,
        \tr\( \hat S(q_0,{\bf q})\tilde\gamma^m  \hat S(q_0-p_0,{\bf q}-{\bf p})\tilde\gamma^n\),
    \label{Pi_expl}
\end{equation}
where the propagator of the quasiparticles in graphene reads
\begin{equation}
    \hat S (q_0,{\bf q}) \equiv \slashed{D}^{-1}\vert_{A_\mu=0} =
        -\frac{(q_0+\mu)\gamma_0-v_F\slashed{q}-m}{(q_0+\mu+i\epsilon {\rm\ sgn}q_0)^2-v_F^2{\bf q}^2-m^2}.
    \label{hat S}
\end{equation}

Note that for $\mu=m=0$ the pole of the propagator yields the linear
dispersion law for quasiparticles in graphene: $q_0 = v_F |{\bf
q}|$. Here $\bq=(q^1,q^2)$, $\slashed{q}=\gamma^1q_1+\gamma^2q_2$.
Vectors with tilde are rescaled by multiplying the spatial
components with $v_F$, i.e., $ \tilde p^j\equiv\eta^j_i p^i=(p_0,v_F
\bp)$, $\eta=\textrm{diag}(1,v_F,v_F)$.

To introduce the temperature in (\ref{Pi_expl}) we perform the
rotation to the Matsubara frequencies
\begin{equation}
 i \int dq_0 \rightarrow - 2 \pi T \sum_{k=-\infty}^{\infty}, \qquad
        q_0  \rightarrow 2\pi i T(k+1/2)  ,
        \label{Mats_prescr}
\end{equation} use the Feynman parametrization
$$
\frac1{ab}=\int_0^1 \frac{dx}{(xa+(1-x)b)^2}
$$
and subsequently change the variables in (\ref{Pi_expl}) in the
spatial part  of the loop--integration:  ${\bf q}\to {\bf q}+x{\bf
p}$. Then we come to
\begin{multline}
 \Pi^{00}=-2 e^2 T N
\sum_{k=-\infty}^{\infty}\int_0^1dx \int\frac{d^2
\bq}{(2\pi)^2}\frac{M_0^2+(q_{0k}+\mu)(q_{0k}+\mu - p_0)}
        {\[(q_{0k}+\mu-xp_0)^2-\Theta^2\]^2} = \\
 - \frac{4\alpha T N}{v_F^2} \sum_{k=-\infty}^{\infty}\int_0^1dx
\int_{\Theta_0}^{+\infty} d\Theta \, \Theta \Biggl(
\frac{\Bigl(\Theta^2 + p_0^2 x(1-x) -2 v_F^2 \bp^2 x(1-x)
\Bigr)}{\[(q_{0k}+\mu-xp_0)^2-\Theta^2\]^2 } + \\
\frac{(q_{0k}+\mu)(q_{0k} + \mu -
p_0)}{\[(q_{0k}+\mu-xp_0)^2-\Theta^2\]^2}\Biggr) \, , \label{Pol00}
\end{multline} where $M_0^2=m^2+ v_F^2\bq^2-x(1-x) v_F^2\bp^2$, and
\begin{align}
    \Theta^2 &=m^2+ v_F^2 \bq^2-x(1-x)(p_0^2-v_F^2 \bp^2) , \\
 \Theta_0 &= \sqrt{m^2-x(1-x)(p_0^2-v_F^2\bp^2)} . \label{th0}
\end{align}
In equation (\ref{Pol00}) we introduced the integration variable
$\Theta$.

In analogy we get
\begin{equation}
\Pi_1^1 + \Pi_2^2 = - 4\alpha T N
\sum_{k=-\infty}^{\infty}\int_0^1dx \int_{\Theta_0}^{+\infty}
d\Theta \, \Theta \, \frac{2m^2 - 2(q_{0k}+\mu)(q_{0k}+\mu -
p_0)}{\[(q_{0k}+\mu-xp_0)^2-\Theta^2\]^2}
\end{equation}

Summation over the fermion Matsubara frequencies can be made
explicitly by making use of the identities \begin{multline}
    \sum_{k=-\infty}^{\infty} \frac1{\[(2 \pi i
    T(k+1/2)-b)^2-\Theta^2\]^2}= \\
        -\frac{1}{16 \Theta^3 T^2}
        \(
        \Theta {\rm\ sech}^2\(\frac{\Theta+b}{2T}\)
        -2 T  \tanh\(\frac{\Theta+b}{2T}\)
        \) +(\Theta\to-\Theta) ,
\end{multline}
\begin{multline}
\sum_{k=-\infty}^{+\infty} \frac{(2 \pi i T(k+1/2) +\mu)(2 \pi i
T(k+1/2) +\mu - p_0) }{\[(2 \pi i T(k+1/2)-b)^2-\Theta^2\]^2} = \\
 -\frac{1}{16 \Theta^3 T^2} \Biggl( {\rm\ sech}^2\(\frac{\Theta+b}{2T}\)
\Bigl(\Theta^3 + \Theta^2 (2x-1)p_0 - \Theta p_0^2 x(1-x) \Bigr) + \\
+  2 T \tanh\(\frac{\Theta+b}{2T}\)\Bigl( \Theta^2 + p_0^2
x(1-x)\Bigr)
  \Biggr)  +(\Theta\to-\Theta), \label{Id2}
\end{multline}
where in (\ref{Id2}) we substituted $b =p_0 x- \mu$.

To perform the integration over $\Theta$ it is convenient to use the
identity $\partial_s \tanh s={\rm sech}^2 s $. Finally we arrive at
the following representation for $\Pi_{00}$ and $ \Pi_{{\rm
tr}}\equiv\Pi_m^m$:
\begin{equation}
  \Pi_{{\rm tr}, 00}
    =-\frac{2 N \alpha T}{v_F^2} \int_0^1dx \(
         f_{tr,00}\tanh\frac{\Theta_0+b}{2 T}
      -
        \ln\(2\cosh\frac{\Theta_0+b}{2 T}\)
        +(\Theta_0\to-\Theta_0) \)
\label{Pi_00} \end{equation} where $\Theta_0\equiv
\sqrt{m^2-x(1-x)(p_0^2-v_F^2{\bf p}^2)}$, $b =p_0 x- \mu$, and
\begin{eqnarray}
    f_{00}&=&\frac{-2 v_F^2{\bf p}^2x(1-x) - p_0(1-2x)\Theta_0+
 2\Theta_0^2}{4 T \Theta_0},
        \label{f_00}\\
    f_{\rm{tr}}&=&\frac{2m^2 v_F^2 + 2 x(1-x) v_F^2  p_3^2}{4 T
    \Theta_0}- \nonumber \\
 & &-\frac{p_0(1-2v_F^2) (1-2x)- 2(1-v_F^2) \Theta_0}{4T}.
        \label{f_tr}
\end{eqnarray}
Here $p_3^2\equiv p_0^2-\bp^2$. We remind that $N$ is the number of
fermion species, $N=4$ for graphene. Parity-odd contributions to the
polarization tensor cancel out between different species, while the
parity-even contributions add up.

\section{Reflection coefficients}

In this section we find reflection coefficients for transverse
electric (TE) and transverse magnetic (TM) modes. The equations
\begin{equation}
\partial_\mu F^{\mu\nu} +\delta(z) \Pi^{\nu\rho}A_\rho =0 \quad
\end{equation}
lead to the conditions
\begin{equation}
\partial_z A_m |_{z=+0} - \partial_z A_m |_{z=-0} = \Pi_{mn} A^n|_{z=0} ,
\label{SQ}
\end{equation}
where \cite{Zeitlin} \begin{equation}
 \Pi^{mn}=\frac1{v_F^2}\eta^m_j \Bigl(
    \Pi^{ji}_0 A(p_0,{\bf p})
    +p_0^2 \Pi^{ji}_u  B(p_0,{\bf p})
    \Bigr) \eta_i^n
    \label{Pi gen tilda}
\end{equation}
\begin{equation}
 \Pi^{ji}_0
    =g^{ji}-\frac{\tilde p^j\tilde p^i}{\tilde p^2},\quad
\Pi^{ji}_u
    =\frac{\tilde p^j\tilde p^i}{\tilde p^2}-\frac{\tilde p^j u^i + u^j \tilde p^i}{(\tilde pu)}
    +\frac{u^ju^i}{(\tilde pu)^2}\tilde p^2 ,
\end{equation}
$u^j=\delta^{j 0} $ and $i,j= 0, 1, 2$. Here $A$, $B$ are scalar
functions.

Let's consider the condition
\begin{equation}
\partial_0 A^0 + \partial_x A^x + \partial_y A^y = 0. \label{gauge}
\end{equation}
In fact, the condition (\ref{gauge}) is quite convenient for a
description of transverse electric and transverse magnetic modes of
the propagating electromagnetic wave.

A nonzero $A_x$, the condition $\partial_x A_x=0$ and the conditions
$A_y=A_z=A_0=0$ describe the propagation of a TE electromagnetic
wave (the electric field is parallel to the surface $z=0$) since
$E_x \sim A_x$. Here the direction perpendicular to the wave vector
of the electromagnetic wave under consideration $(0, p_y, p_z)$ is
denoted by $x$.

For the TE wave we have:
\begin{align}
A_x &= e^{i p_y y} e^{i p_z z} + r_{TE} e^{i p_y y}e^{-i p_z z}
\quad \text{for} \quad z<0  \label{TE11}
\\ A_x &= e^{i p_z z} e^{i p_y y} t_{TE} \quad \text{for} \quad z>0 \label{TE2}
\end{align}
and
\begin{equation}
\Pi_{xn} A^n = A_x A(p). \label{T1}
\end{equation}
Here $p_z^2=p_0^2-p_y^2$. From the continuity of potentials at $z=0$
we obtain $1+r_{TE}=t_{TE}$. Now one substitutes (\ref{TE11}) and
(\ref{TE2}) into (\ref{SQ}) and uses (\ref{T1}) to obtain:
\begin{equation}
r_{TE} = \frac{A(p)}{2ip_z - A(p)} \label{rte2}
\end{equation}

The conditions $A_x=A_z=0$, $p_0 A_0=p_y A_y$ describe the
transverse magnetic (TM) wave. This choice of vector potentials
describes TM wave since $E_z \sim
\partial_z A_0 $ or $B_x \sim \partial_z A_y$.
For  $A_0$ we have:
\begin{align}
A_0 &= e^{i p_y y} e^{i p_z z} + r_{A_0} e^{i p_y y}e^{-i p_z z}
\quad \text{for} \quad z<0  \label{TM1}
\\ A_0 &= e^{i p_z z}e^{i p_y y} t_{A_0} \quad \text{for} \quad z>0 \label{TM2}.
\end{align}
and
\begin{equation}
\Pi_{0n} A^n = \bigl(A(p) - B(p) v_F^2 p_y^2 \bigr) p_z^2/\tilde
p_z^2 A_0 ,\label{T3}
\end{equation}
where $\tilde p_z^2= p_0^2- v_F^2 p_y^2$. One substitutes
(\ref{TM1}) and (\ref{TM2}) into (\ref{SQ}) and uses (\ref{T3}) to
obtain:
\begin{equation}
r_{A_0} = \frac{ p_z(A(p) - p_y^2 v_F^2 B(p)) }{ 2 i \tilde p_z^2 -
p_z (A(p) - p_y^2 v_F^2 B(p)) }
\end{equation}
Since $E_z \sim \partial_z A_0$ the reflection coefficient for the
TM mode is equal
\begin{equation}
r_{TM}= - r_{A_0} = -\frac{ p_z(A(p) - p_y^2 v_F^2 B(p)) }{ 2 i
\tilde p_z^2 -  p_z (A(p) - p_y^2 v_F^2 B(p)) } \label{rtm2}
\end{equation}

The reflection coefficients (\ref{rte2}) and (\ref{rtm2}) are
reflection coefficients of transverse electric and transverse
magnetic modes respectively.

One can also rewrite the reflection
  coefficients in terms of the polarization tensor components
\begin{equation} r_{\rm TM}=\frac{p_z  \Pi_{00}}{p_z \Pi_{00}  + 2 i p_y^2},
\qquad r_{\rm TE}= - \frac{ p_z^2 \Pi_{00}+ p_y^2 \Pi_{\rm tr}}
            {p_z^2 \Pi_{00} +  p_y^2 (\Pi_{\rm tr} - 2 i p_z)}.
\label{rTETM-grPi}
\end{equation}

\section{QED point of view}

 The two conditions follow from gauge
invariance:
\begin{align}
p_0 \Pi_{00} (p_0) - p_y \Pi_{y0} (p_0) &= 0 ,\nonumber\\
p_0 \Pi_{0y} (p_0) - p_y \Pi_{yy} (p_0) &= 0 , \nonumber
\end{align}
which yield after Wick rotation
\begin{equation}
p_0^2 \Pi_{00}(i p_0) = - p_y^2 \Pi_{yy} (i p_0) \label{ID1}
\end{equation}
 and the property
\begin{equation}
\Pi_{\rm tr} (i p_0) = \Pi_{00} (i p_0) \frac{p_0^2 + p_y^2}{p_y^2}
- \Pi_{xx} (i p_0).
\end{equation}


 The reflection coefficients can be rewritten in the form:
\begin{align}
r_{\rm TM}(i p_0) &=  - \frac{\sqrt{p_0^2 + p_y^2} \Pi_{yy}(i
p_0)}{2 p_0^2} \Biggl(1 -  \frac{\sqrt{p_0^2+ p_y^2} \Pi_{yy}(i
p_0)}{2 p_0^2} \Biggr)^{-1},  \label{Par1} \\
 r_{\rm TE}(i p_0) &= \frac{\Pi_{xx}(i
p_0)}{2\sqrt{p_0^2+ p_y^2}} \Biggl(1- \frac{\Pi_{xx}(i
p_0)}{2\sqrt{p_0^2+ p_y^2}}\Biggr)^{-1} .\label{Par2}
\end{align}
 In the gauge $A_0=0$ the longitudinal part of the free photon
propagator $D^L$ has the form
\begin{equation}
D^L (ip_0) =  \frac{ \sqrt{p_0^2 + p_y^2} e^{-|z|\sqrt{p_0^2 +
p_y^2}} }{2 p_0^2} , \label{DL}
\end{equation}
the transverse part of the free photon propagator $D^T$ has the
form:
\begin{equation}
D^{T} (ip_0) = \frac{e^{-|z|\sqrt{p_0^2+ p_y^2}}}{2\sqrt{p_0^2+
p_y^2}} . \label{DT}
\end{equation}

Let's choose coordinate axes in the plane of a graphene sheet so
that $p_y = {|\bf p|}$. Lifshitz free energy \cite{Lifshitz} has the
form:
\begin{equation}
    {\mathcal F}
    =T\sum_{n=-\infty}^\infty\int\frac{d^2{\bf p}}{8\pi^2} \ln [(1-e^{-2a\sqrt{\omega_n^2+{\bf p}^2}}r_{\rm
 \rm TE}^{(1)}r_{\rm \rm TE}^{(2)})
        (1-e^{-2a\sqrt{\omega_n^2+{\bf p}^2}}r_{\rm \rm TM}^{(1)}r_{\rm \rm TM}^{(2)})] ,
        \label{EL}
\end{equation}
where $\omega_n=2\pi n T$ are Matsubara frequencies, the respective
transverse magnetic and transverse electric reflection coefficients
from two  parallel flat surfaces separated by a vacuum slit $a$ are
denoted by $r_{TM}$ and $r_{TE}$. For the ideal metal $r_{TM}=1$,
$r_{TE}=-1$. Note, however, that some exact results in complicated
geometries\cite{Mar2}-\cite{Mar3} can be essentially different from
the approximations based on Lifshitz formula for two parallel plates
(see also Ref.\cite{Pirozhenko} which considers sphere-plane
system).

From the comparison of the formula (\ref{EL}) and expressions
(\ref{Par1}) -- (\ref{DT}) it follows that for graphene -- ideal
metal, graphene -- graphene the  Lifshitz theory takes into account
the set of closed Feynman one loop diagrams responsible for
interaction between the two materials separated by a vacuum slit.
The longitudinal and the transverse parts of the photon propagator
enter two different sets of closed one loop diagrams for the free
energy and multiplied in these Feynman diagrams by the longitudinal
and the transverse components of the polarization operator
$\Pi_{yy}$ and $\Pi_{xx}$ respectively.
 Lifshitz type formulas result from the
sum over closed Feynman one loop diagrams with $z=a$ or $z=2a$ in
photon propagators connecting the two sheets of graphene or a
graphene sheet interacting with an ideal metal respectively. Inside
each graphene layer the sum of RPA diagrams is taken into account by
factors in round parentheses in (\ref{Par1}), (\ref{Par2}). The
division of the free energy into longitudinal and transverse parts
in terms of respective parts of photon Green's functions and
polarization operator is equivalent to a division into TM and TE
parts described by the reflection coefficients $r_{\rm TM}$ and
$r_{\rm TE}$ in the Lifshitz approach.

\section{High-temperature asymptotics}

 Let's assume $m=0$, $\mu=0$
 \cite{Mahan}.
For $|{\bf p}| \to 0$ one gets:
\begin{align}
&\Pi^{00}(ip_0=0) =  \frac{4 \alpha N T \ln2}{v_F^2}
+\frac{\alpha N {\bf p}^2}{12 T} + \dots ,  \nonumber \\
 &{\rm tr}\Pi(i p_0=0) - \Pi^{00} (i p_0=0) =
  \frac{\alpha N v_F^2 {\bf p}^2}{6T} + \dots , \nonumber
\end{align}
and reflection coefficients have the form:
\begin{align}
r_{TE}(ip_0=0) &\mathop{\simeq} -\frac{\alpha N v_F^2 |{\bf p}|}{\alpha N v_F^2 {|{\bf p}|} + 12 T} , \nonumber\\
 r_{TM}(ip_0=0) &\mathop{\simeq} \frac{2\alpha N T \ln2
 + \alpha N {\bf p}^2 v_F^2/(24 T)}{
2\alpha N T\ln2 + \alpha N {\bf p}^2 v_F^2/(24 T) + |{\bf p}| v_F^2}
.\nonumber
\end{align}

Zero Matsubara TM and TE terms yield the following high-temperature
behavior of the free energy (\ref{EL}) in graphene -- ideal metal
system:
\begin{align}
  {\mathcal{F}}_{0 \rm TM } &=
       -\frac{T\zeta(3)}{16 \pi a^2} + \frac{v_F^2 \zeta(3)}{32\alpha
       N\pi (\ln2)a^3} + \dots
               ,  \\
  {\mathcal F}_{0\rm TE}
         &=  -\frac{\a N v_F^2 }{192 \pi a^3} + \dots.
\end{align}
Here
\begin{equation}
-\frac{T\zeta(3)}{16 \pi a^2} \equiv {\mathcal F}_{\rm
Drude}\vert_{T\to\infty}=\frac12 {\mathcal F}_{\rm
id}\vert_{T\to\infty} .
\end{equation}
is the high-temperature  asymptotics of the metal -- metal system
with a Drude model of the permittivity used
\cite{Sernelius2}-\cite{Drude1}, which is equal to one half of the
high-temperature asymptotics in the metal -- metal system with the
ideal boundary conditions or the plasma model of the permittivity
used \cite{Decca} (see Ref.\cite{Brevik7} for a review). The zero
frequency TE Matsubara term is suppressed by a factor $\alpha N
v_F^2$ and additional power of $1/(Ta)$.

The typical region of validity of the high-temperature asymptotics
for the metal--metal system is $4\pi T a \gg 1$. However, due to a
suppression of nonzero Matsubara terms by the coupling constant
$\alpha$ the free energy in a graphene -- metal system approaches
the high-temperature asymptotics $-T\zeta(3)/(16\pi a^2)$ much
quicker and at shorter separations than in the metal -- metal case
(see next section).

Note that in the graphene -- ideal metal system the high-temperature
asymptotics is derived from the first principles of quantum field
theory.

\section{Nonzero Matsubara terms}
It is often desirable to have an accurate analytical approximation
of the exact result at different separations. We present such an
expression for the sum of nonzero Matsubara terms in this section.

To obtain an appropriate analytical expression we first note that at
separations $H\gg v_F$ one can put $v_F=0$ in any nonzero Matsubara
term. It is possible due to the exponential factor in the Lifshitz
formula which effectively restrains the integration over impulse to
$a |{\bf p}|\lesssim1$. In this case contribution of the type of
$v_F^2 (a |{\bf p}|)^2$ can be neglected compared to $(ap_0)^2=(2
\pi n aT )^2$ due to the smallness of the parameter $v_F$.

In the finite temperature sum of nonzero Matsubara terms in the
Lifshitz formula one can use the reflection coefficients taken at
zero temperature. The corrections due to finite temperature are
suppressed for nonzero Matsubara terms, so we neglect them in the
leading approximation.

Under two mentioned above approximations and the condition $m=\mu=0$
the reflection coefficients of a single graphene layer at zero
temperature have the form:
\begin{align}
r_{\rm TM}^{T=0} &= \frac{\pi \alpha N \sqrt{p_0^2 + {\bf p}^2} }{
\pi
    \alpha N \sqrt{p_0^2 + {\bf p}^2} + 8 \sqrt{p_0^2 + v_F^2 {\bf p}^2} }
    \simeq \frac{\pi \alpha N \sqrt{p_0^2 + {\bf p}^2} }{ \pi
    \alpha N \sqrt{p_0^2 + {\bf p}^2} + 8 |p_0|}
    \label{rtm}\\
r_{\rm TE}^{T=0} &=  - \frac{\pi \alpha N \sqrt{p_0^2 + v_F^2 {\bf
p}^2} }{
    \pi \alpha N \sqrt{p_0^2 + v_F^2 {\bf p}^2} + 8 \sqrt{p_0^2 + {\bf p}^2} }
    \simeq  - \frac{\pi \alpha N |p_0|}
        {\pi \alpha N |p_0| + 8 \sqrt{p_0^2 + {\bf p}^2} }
    \label{rte}
\end{align}

\begin{figure}
\centering \includegraphics[width=12cm]{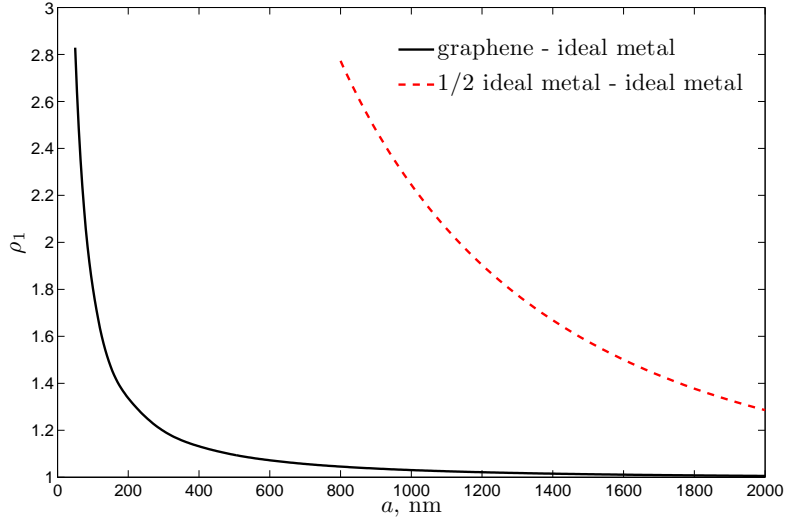} \caption{Ratio
$\rho_1$ of the free energy to the leading high-temperature
asymptotics $-T\zeta(3)/(16\pi a^2)$. Both graphs are evaluated for
$T=300$K. In graphene the values $m=\mu=0$ were used.}
\label{Mayplot}
\end{figure}

Due to smallness of the reflection coefficients (both being of the
order of $\alpha$) we can take just the first term in the expansion
of the logarithm in the Lifshitz formula. Note, however, that
expansion of the TM reflection coefficient in $\alpha$ is not
legitimate, as will become evident below (see (\ref{appt})).

\begin{figure}
\centering \includegraphics[width=12cm]{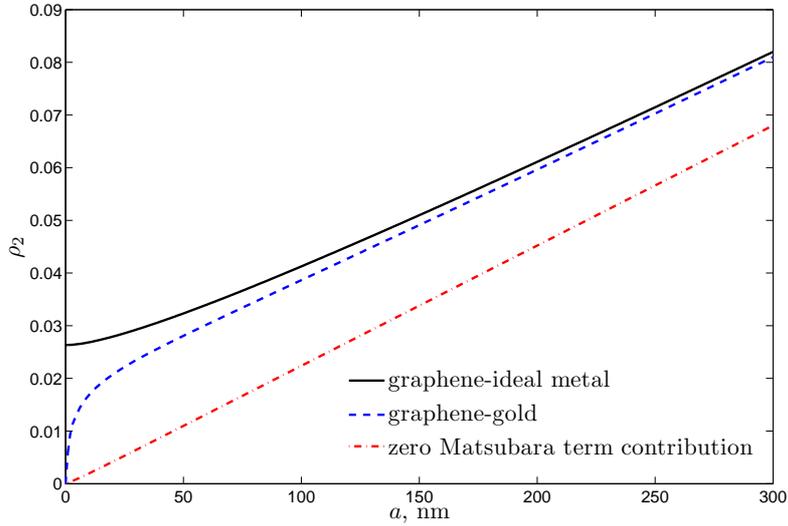} \caption{Ratio
$\rho_2$ of the free energy for a graphene - metal system with
$\mu=m=0$ to the ideal metal - ideal metal free energy at $T=300$K.}
\label{ratio1}
\end{figure}

The sum of nonzero Matsubara terms in (\ref{EL}) in this
approximation in the TM case with $r^{(1)}_{\rm TM} = r_{\rm
TM}^{T=0}$ (\ref{rtm}) and $r^{(2)}_{\rm TM}=1$ equals to
\begin{align}
\Delta\mathcal{F}_{\rm TM} &= - \frac{T}{2\pi} \sum_{n=1}^{+\infty}
    \int_{Hn/2}^{+\infty} ds_1 \frac{s_1^2}{s_1+ 16nT/(\alpha N)}
    \exp(-2as_1) = \nonumber \\
    &= - \frac{T}{8\pi a^2} \sum_{n=1}^{+\infty} \exp(-Hn) (1 - gn + Hn)
    + (gn)^2 \exp(gn) {\rm E_1}(gn + Hn) , \label{TMint2}
\end{align}
here $s_1=\sqrt{\omega_n^2+ |{\bf p}|^2}$ and $g \equiv 32 T
a/(\alpha N)$, $E_1$ stands for the standard exponential integral
function. It is convenient to reexpress the result (\ref{TMint2}) in
an integral form. For this purpose one has to differentiate
(\ref{TMint2}) over $H$, assuming $H$ as an independent parameter
for the moment, calculate the sum over $n$ and then integrate back
over $H$ (the integration constant is fixed as zero at
$H\to\infty$). Thus one obtains
\begin{equation}
\Delta\mathcal{F}_{\rm TM} = -\frac{T \alpha N}{8 a^2}
    \int_{H}^{+\infty} dt \frac{\exp(t)\bigl(\exp(t)+1 \bigr)\:
    t^2}{\bigl(\exp(t)-1 \bigr)^3 \bigl(8H + \pi\alpha N t \bigr)}
    \label{Intrepr}
\end{equation}

The TE part of the nonzero Matsubara terms of the Lifshitz formula 
with the coefficients $r^{(1)}_{TE} = r_{TE}^{T=0}$ from (\ref{rte})
and $r^{(2)}_{TE}= -1$   gives  the following contribution in the
leading order in $\a$
\begin{equation}
\Delta\mathcal{F}_{TE} \simeq - \frac{T^2 \pi \alpha N }{8}
    \sum_{n=1}^{+\infty} n \int_{\omega_n}^{+\infty} ds_1 \: \exp(-2 a s_1)
    = - \frac{T^2 \pi \alpha N }{16 a} \frac{\exp(-H)}{(1-\exp(-H))^2} .
    \label{TE1}
\end{equation}

Thus, the complete result for the sum of nonzero Matsubara TM and TE
terms in the approximation described above is given by
(\ref{Intrepr}) and  (\ref{TE1}):
\begin{equation}
    \Delta\mathcal{F}  = \Delta\mathcal{F}_{TM} +\Delta\mathcal{F}_{TE}.
    \label{nonz}
\end{equation}
Consequently, the leading $v_F=0$ contribution to the free energy is
the sum $-T\zeta(3)/(16\pi a^2)$ + $\Delta \mathcal{F}$. It can be
used for the comparison of the theory and experiment with $1\%$
accuracy for all separations at $T=300$K.

From (\ref{Intrepr}) and (\ref{TE1}) one gets the energy at $T=0$ in
the limit $v_F\to 0$:
\begin{equation}
\Delta\mathcal{F}\bigl|_{T\to 0} =
    -\frac{\alpha N}{128\pi a^3} \ln\bigl(1+ 8/(\alpha N \pi)\bigr)
    - \frac{\alpha N}{256 \pi a^3 }  , \label{appt}
\end{equation}
where the non analyticity in $\alpha$ comes from the TM mode.


Fig.\ref{Mayplot} clearly demonstrates that the free energy of a
graphene -- metal system approaches the high-temperature asymptotics
$-T\zeta(3)/(16\pi a^2)$ much quicker and at shorter separations
than the two metals' system. Such an approach to the
high-temperature behavior is related to the fact that nonzero
Matsubara terms are of the order of the coupling constant $\alpha$
and thus very small in comparison with respective terms for metals.
 The zero frequency TM
Matsubara term acquires the value of the sum of nonzero Matsubara
terms in the free energy at separations $a\approx 100$nm at $T=300$K
(see Fig.\ref{ratio1}) and dominates in the free energy at larger
separations.  Thus the high-temperature behavior in graphene --
metal systems should be observed at separations of the order of
$100$ nm at $T=300$K (the same effect in metal -- metal systems
takes place at separations of the order of several micrometers at
$T=300$K).

\section{Conclusions}

The behavior of the free energy of the graphene -- metal system is
studied on the basis of the field theoretic model. The components of
the polarization operator of $2+1$ quasiparticles in a graphene
layer are evaluated at finite temperature.  The TM and TE reflection
coefficients are derived from the solutions of the boundary problems
for vector potentials.

In the high-temperature limit the asymptotics of the free energy
coincides with the Drude model asymptotics for two metals' system.

The crossover to the high-temperature behavior in a graphene-metal
system takes place at separations $a$ of the order of $100$ nm at
$T=300$K. This is the reason why the systems with graphene are very
promising for the experimental studies of the finite-temperature
Casimir effect.

\section*{Acknowledgments}
The author is grateful to the organizers of QFEXT-11 for support.
The author is grateful to colleagues for numerous discussions in
Benasque.


\end{document}